\documentclass[twocolumn,showpacs,preprintnumbers,amsmath,amssymb]{revtex4}

\usepackage{graphicx}
\usepackage{dcolumn}
\usepackage{bm}

\begin{document}

\title{Impact of densitized lapse slicings on evolutions of a wobbling black hole}

\author{Ulrich Sperhake}
\affiliation{Centers for Gravitational Physics \& Geometry and Gravitational
             Wave Physics, \\
             Penn State University, University Park, PA 16802}
\author{Kenneth L. Smith}
\affiliation{Centers for Gravitational Physics \& Geometry and Gravitational
             Wave Physics, \\
             Penn State University, University Park, PA 16802}
\author{Bernard Kelly}
\affiliation{Centers for Gravitational Physics \& Geometry and Gravitational
             Wave Physics, \\
             Penn State University, University Park, PA 16802}
\author{Pablo Laguna}
\affiliation{Centers for Gravitational Physics \& Geometry and Gravitational
             Wave Physics, \\
             Penn State University, University Park, PA 16802}
\author{Deirdre Shoemaker}
\affiliation{Center for Radiophysics and Space Research\\
             Cornell University, Ithaca, NY 14853}

\date{\today}

\begin{abstract}
We present long-term stable and second-order convergent
evolutions of an excised wobbling black hole. Our results 
clearly demonstrate that
the use of a densitized lapse function 
extends the lifetime of simulations dramatically. We also show the improvement 
in the stability of single static black holes when an algebraic densitized 
lapse condition is applied. In addition, 
we introduce a computationally inexpensive approach for 
tracking the location of the singularity suitable for mildly distorted
black holes. The method is based on
investigating the fall-off behavior and asymmetry of appropriate grid variables.
This simple tracking method allows one 
to adjust the location of the excision region to follow the coordinate
motion of the singularity.
\end{abstract}

\pacs{04.25.Dm}

\maketitle

\section{Introduction}
The accurate, long-term stable evolution of binary black hole spacetimes
is one of the most important challenges in general relativity
research. While the solution of the relativistic two-body problem
is highly relevant from an academic point of view, its importance is
further enhanced by the need of such simulations for predicting the
resulting gravitational wave emission. This will not only improve the
chances of detecting gravitational waves but also be essential when it
comes to the astrophysical interpretation of the measured wave patterns. 

Early black hole evolutions (see e.g.\,\cite{Anninos1995b, Cook1998,
Brandt2000}) were based on the canonical
ADM formulation \cite{Arnowitt1962,York1979}
of the Einstein field equations. The stability properties of these
evolutions, however, were severely limited. It gradually
became accepted among the community that the underlying problems were
not merely numerical in nature, but originated at least in part from
the structure of the equations themselves. The striking success achieved in
characteristic evolutions of single black hole spacetimes
\cite{Gomez1998,Gomez1998b} supports this view. As a result, numerous
alternative
3+1 formulations have been constructed in recent years. The majority
of these formulations are modified versions of the original ADM scheme. These
alternative formulations either recast the Einstein field equations
in explicitly hyperbolic form (see \cite{Kidder2001} and references therein)
or exhibit some ``hyperbolic
flavor'' such as the Baumgarte-Shapiro-Shibata-Nakamura (BSSN) system
\cite{Shibata1995,Baumgarte1999}. 

In recent years, several of these
new systems have demonstrated vastly improved stability
properties, either in fully non-linear 3-dimensional black hole evolutions
or suitably chosen test problems which isolate some of the particular
difficulties faced in numerical relativity. Even though the question
as to the optimal formulation of the Einstein equations remains as yet
unanswered, the development of alternative 3+1 schemes has been one of the
key driving forces underlying recent progress in numerical black hole
evolutions. 

In addition to the form used to recast the Einstein equations,
considerable improvements have also been achieved in dealing with
spacetime singularities inevitably present in such simulations. One approach,
commonly known as {\em singularity avoidance}, foliates the spacetime in such a way
that the evolution of the
code is basically frozen near the singularity. As originally implemented,
singularity avoidance methods enjoyed rather limited success,
mainly due to the `grid-stretching' effect --- the increase of the
proper separation between neighboring grid-points. The problem
of grid-stretching
has been alleviated to some extent by 
the introduction of a non-zero shift vector~\cite{Alcubierre2003b}. 
It is not clear, however, whether
this approach will be able to sustain evolutions for the dynamical time-scales 
that are needed. 
   
An alternative treatment,
often attributed to a suggestion by Unruh (quoted in
\cite{Thornburg1987}), is based
on the cosmic censorship conjecture. Cosmic censorship states that a
spacetime singularity will be surrounded by a causal boundary, the black hole horizon.
Physical information contained within this boundary cannot escape. 
Thus, in principle, one can get away with murder and excise
a region inside the event horizon containing the black hole singularity, 
without affecting computations in the exterior. 
This technique, known as {\em black hole excision} or the
apparent horizon boundary condition, has become very popular in
recent years and has been implemented in a number of cases
\cite{Seidel1992, Anninos1995, Marsa1996, Scheel1997, Cook1998, Gomez1998,
Brandt2000, Alcubierre2001, Yo2001, Alcubierre2001b, Shoemaker2003,
Calabrese2003}.
It is important to note, however, that the causal boundary
strictly applies to physical information only. Unless one can guarantee
that all characteristic speeds of the evolved fields
(physical and unphysical) lie within the light cone (for example
by using an explicitly hyperbolic formulation),
gauge modes may well leave the black hole interior.
While these modes are not supposed to affect the physical results in the
exterior evolution, they may have a significant impact on the numerical
stability. 
The success of the excision technique as applied to
non-manifestly-hyperbolic
systems of equations is therefore by no means a foregone conclusion and
needs to be demonstrated explicitly for the application at hand. 

Earlier implementations of black hole excision were still based on the ADM
scheme and frequently made use of causal-differencing and/or apparent
horizon locking coordinates \cite{Cook1998, Seidel1992, Anninos1995}.
Aside from demonstrating the feasibility of the method, these simulations
led to improved evolutions of single black hole spacetimes as
compared with techniques based on singularity
avoiding slicing conditions. These codes were, however, still subject to the
fundamental limitations of the ADM formulation. In more recent times,
excision has been combined with alternative formulations
of the field equations and more general classes of gauge conditions.
The resulting evolutions have been able to reach the
first major goal on the way toward solving the binary black hole problem,
namely the evolution of single, stationary black hole spacetimes
in three dimensions for arbitrary times \cite{Alcubierre2001b, Yo2002, Gomez1998}. 
In addressing dynamic black hole scenarios,
one may attempt to absorb part or all of the black hole movement
by choosing appropriate coordinates
such as co-rotating coordinates in the case
of an in-spiraling binary black hole \cite{Brady1998}.
However, it is unclear how
such a treatment will be possible, in particular in the later stages
just before the merger, without introducing a considerable
degree of grid-stretching.
It is unlikely, therefore, that such dynamic simulations can be
achieved without an excision region that can adjust
dynamically to mask the black hole singularity as it moves through the
computational domain.
In the words of Ref.~\cite{Scheel1997}, ``a key milestone in
three-dimensional black hole simulations is the ability to stably move a hole through
the numerical grid''.
 
In characteristic formulations, long-term stable evolutions using a dynamic
excision region have been achieved as early as 1998
for the case of a wobbling black hole \cite{Gomez1998b}.
In 3+1 formulations, to our knowledge,
the earliest attempts were performed by \citet{Cook1998}, who managed to track
a boosted
black hole through the grid for 60\,$M$ ( where $M$ is the mass of the 
black hole). Further applications
of dynamic singularity excision include grazing collisions of black holes
\cite{Brandt2000} and the evolution of a scalar field in a static or boosted
Kerr-Schild black hole background \cite{Yo2001}. In \cite{Shoemaker2003},
a dynamic scenario was simulated by evolving a single black hole in
ingoing-Eddington-Finkelstein (iEF) coordinates transformed to a
coordinate system in which the coordinate location of the singularity traced out
bouncing or circular paths on the coordinate grid. 
As the excision region
is moved across the numerical grid in such evolutions, one will
inevitably encounter previously excised grid points which reemerge
into the evolved domain and thus need to be filled with valid data
for the grid variables. In evolutions with dynamic excision, this is
commonly done via extrapolation from neighboring grid points.
The motivation for this procedure arises from the same
causal considerations underlying the singularity excision
mentioned above and is subject to the same reservations for
non-manifestly-hyperbolic
evolution systems. Several extrapolation techniques have
been studied in \cite{Yo2001}, where the stability properties
of the ensuing evolutions are found to be rather insensitive to the precise
details of the scheme. 

In this paper, we expand directly on the work of \cite{Shoemaker2003}
(referred to as Paper I from now on) where third-order
extrapolation was used for the population of reemerging grid points.
The resulting dynamic excision enabled these authors to evolve
non-boosted black holes traveling on circular or bouncing trajectories
for about $130\,M$. While such evolution times represent considerable
progress on previous dynamic simulations, they do not meet the requirements
for simulations of in-spiraling binary black holes. There remains, thus,
the fundamental question of whether the dynamic excision techniques
currently under consideration
are in principle capable of handling long-term stable evolutions of moving
black holes. Addressing this question is one of the main goals
of this paper. We demonstrate
how the lifetime of the dancing or wobbling black hole evolutions of Paper I
can be extended to at least several thousands of $M$ by prescribing
the slicing condition in terms of the densitized lapse as opposed to the
lapse function itself. This also confirms the belief expressed in Paper I
that the gauge conditions were a crucial factor in limiting the
stability properties of the code. 
We further present a simple and very efficient mechanism by which
our excision region was able to track the black hole movement
by merely analyzing the updated numerical
data, with no prior knowledge of the black hole trajectory. This
mechanism analyzes the fall-off behavior of certain grid variables
near the black hole singularity and
provides us with an alternative at negligible computational cost
to more commonly used apparent horizon tracking methods. 
This method for tracking singularities is suitable for
situations in which the black hole is mildly distorted.

The numerical code used for the evolutions presented in this paper is the
Maya code; details of the code can be found in Paper I. The code is based
on the BSSN formulation \cite{Shibata1995, Baumgarte1999}; we will not
repeat the BSSN equations here --- see e.g.\,\cite{Alcubierre2003b}.
Whenever we encounter conformally transformed quantities, we will
distinguish them by a ``hat'' from physical quantities.
Latin and Greek indices will run from 1 to 3 and 0 to 3 respectively.
We define the order of an extrapolation scheme
as the order of the polynomial used. 

The paper is organized as follows.
We describe and motivate in Sec.~\ref{DENSITIZEDLAPSE}
the use of a densitized lapse for our numerical evolution. The numerical
results together with convergence tests will be presented in Sec.~\ref{RESULTS}.
In Sec.~\ref{GAUSSTRACKING}, we describe the simple scheme
used in our evolutions for tracking the black hole singularity.
We conclude in Sec.~\ref{CONCLUSIONS} with a discussion of our results.

\section{Introducing a densitized lapse}
\label{DENSITIZEDLAPSE}
Under the canonical 3+1 decomposition of the Einstein equations,
the line element is given by
\begin{equation}
  ds^2 = -\alpha^2dt^2 +\gamma_{ij} (dx^i + \beta^i dt)(dx^j + \beta^j dt),
         \label{ADMLINEELEMENT}
\end{equation}
where the 3-metric $\gamma_{ij}$ describes the geometry of the space-like
slices. The lapse function $\alpha$ relates the separation in proper time
between two space-like slices to their difference $dt$ in coordinate time,
and the shift vector $\beta^i$ determines how one identifies points with
identical spatial coordinates on neighboring spatial hypersurfaces.
For recasting the Einstein equations in a 3+1 form,
it is convenient to also introduce the extrinsic curvature
\begin{equation}
  K_{ij} = -\frac{1}{2\alpha} (\partial_t - \mathcal{L}_{\beta}) \gamma_{ij},
\end{equation}
where $\mathcal{L_{\beta}}$ denotes the Lie derivative along the shift vector.
The extrinsic curvature describes the embedding of the 3-dimensional spatial
slices in 4-dimensional spacetime.
The 3+1 Einstein equations fall into two categories,
12 evolution equations for $\gamma_{ij}$ and $K_{ij}$, and 4 constraint
equations (the Hamiltonian and momentum constraints).
The majority of formulations of the Einstein field equations,
and in particular the BSSN formulation we use for our code, are based
on this decomposition but introduce various modifications, such
as conformal decompositions,
promoting auxiliary variables to fundamental status and combining the
evolution and constraint equations to obtain systems with different
mathematical and numerical properties.
 
The gauge functions $\alpha$ and $\beta^i$, on the other
hand, represent the coordinate or gauge freedom of general relativity
and can be chosen arbitrarily without affecting the physical spacetime.
Their choice does, however, have a crucial impact on the performance
of a numerical scheme used to evolve the Einstein equations.
Since the evolution of the gauge variables is not determined by the
Einstein equations, one needs to introduce a recipe to determine
the the lapse and shift. Such gauge conditions can
in principle be divided into three types
(see for example \cite{Khokhlov2002}). {\em Fixed} gauge conditions
are those for which $\alpha$ and $\beta^i$ are given functions of the spacetime coordinates
$x^{\alpha}$. {\em Algebraic} conditions, on the other hand, are such that 
the lapse and shift are allowed to also depend
on other evolved variables, such as $\gamma$, $K$ or derivatives of
$\gamma_{ij}$. Finally, one may construct
$\alpha$ and $\beta$ from solutions of differential equations (elliptic,
parabolic or hyperbolic). This type is sometimes referred to as {\em differential}
gauge conditions. Examples are minimal distortion shift and maximal slicing. 
A subclass of algebraic gauge conditions will be 
of particular importance for this work. 

For the present work, we will only consider a fixed gauge condition for
the shift vector; that is, the shift vector will be given by the exact solution
for a single, non-rotating black hole in iEF
coordinates. As mentioned before, in Paper I we also used a fixed gauge
condition for the lapse function. The objective now is to
replace this with an algebraic condition.
In particular the algebraic lapse we consider is
\begin{equation}
  \alpha = \gamma^{n/2}\,q = e^{6n\phi}\,q, \label{DEFQ}
\end{equation}
where $\phi$ is the conformal factor relating the physical metric
$\gamma_{ij}$ to the conformal metric $\hat\gamma_{ij}$ such that
\begin{equation}
\gamma_{ij} = \gamma^{1/3}\hat\gamma_{ij} = e^{4\phi}\,\hat\gamma_{ij}\,.
\end{equation} 
We emphasize that in Eq.~(\ref{DEFQ}), $q$ replaces the lapse $\alpha$
as the independent variable. Notice that
because $\alpha$ is a true scalar, $q$ is a density of weight $-n$, 
hence the name densitized lapse.

Such a densitized lapse or a generalized version thereof has played a
crucial rule in the
development of various hyperbolic formulations of Einstein's field
equations. For example Kidder et al.\,\cite{Kidder2001} find the use of a
densitized version of the lapse
necessary for obtaining strong hyperbolicity. In particular these authors
obtain exclusively physical characteristic speeds for the case corresponding
to $n=1$ in our Eq.\,(\ref{DEFQ}). Similarly, Frittelli and Reula
\cite{Frittelli1996} require
a densitized lapse with $n>0$ in order to obtain a symmetric hyperbolic
system.
In Ref.~\cite{Calabrese2002} 1-dimensional non-linear evolutions are presented
to highlight the markedly different convergence and stability properties
of strongly hyperbolic, weakly hyperbolic and completely ill-posed
problems.
Note that these authors switch between the different schemes
by describing the spacetime slicing in terms of a densitized lapse
of weight $n=+1,\,\,0,\,\,-1$ (translated into our notation). Our results
are compatible with their findings in that we observe extremely poor
behavior in evolutions using $n=-1$, somewhat better --- though not convincing ---
behavior with $n=0$, and (as is demonstrated below) long-term
stability and convergence for dynamic scenarios in three dimensions
on the same time scale as in their evolutions by using $n=+1$.

With respect to the BSSN system used in our calculations, it is of
particular interest to note that Sarbach et al.\,\cite{Sarbach2002}
managed to establish equivalence between the BSSN system and hyperbolic
formulations under certain conditions. One of these conditions is again the use
of a densitized lapse with $n>0$. They also find that the special
case $n=1$ leads
to exclusively physical characteristic speeds. In a more general analysis of the
stability of gauge conditions carried out within the framework of
short wavelength perturbations, Khokhlov et al.\,\cite{Khokhlov2002}
demonstrate that the use of fixed gauge conditions
yields ill-posedness in all cases except for
vanishing shift and spatially constant lapse. Allowing the lapse to be
a function of coordinates and the volume element
$\gamma$ on the other hand
they obtain a well-posed system provided that $\partial \alpha^2
/\partial \gamma > 0$. For the restricted class of densitized lapse
defined in Eq.\,(\ref{DEFQ}) this is equivalent to the requirement
that $n>0$. It is also interesting to note that hyperbolic systems
have resulted in significantly longer evolution times for a stationary
single black
hole than comparable evolutions of the BSSN or ADM systems
if the gauge was specified in the form of given functions
of $x^{\alpha}$. Scheel at al. \cite{Scheel2002}, for example, present
lifetimes of 3-dimensional black hole evolutions about two orders of
magnitude longer than those commonly achieved with non-hyperbolic
systems where the lapse (as opposed to the densitized lapse) has
been specified algebraically. To our knowledge,
the only
case where a densitized version of the lapse has been used in combination
with a BSSN-like system is that of Laguna and Shoemaker
\cite{Laguna2002}, who also report
substantially improved stability properties. The results presented
below will indeed strongly indicate that the use of a densitized lapse
is the key ingredient for obtaining the improved stability properties
in these cases.
 
As mentioned before, in our numerical evolutions we will mainly focus
on an algebraic slicing which
requires us to prescribe $q$ as a given function of time and space and
calculate the lapse $\alpha$ from Eq.\,(\ref{DEFQ}). Because
$\phi$ in Eq.~(\ref{DEFQ}) is a numerically evolved quantity,
in a qualitative manner of speaking, we
enable the gauge conditions to respond in a more flexible way to
the numerical evolution. 

For our evolutions of static black holes, we will also consider the case
of a differential slicing condition.
In particular we choose the live 1+log slicing
which has been used with great success in such scenarios
\cite{Yo2002, Alcubierre2001b}. This slicing condition is a special case
of the so-called modified Bona-Mass{\'o} slicing
\begin{equation}
  \partial_t\alpha = -\alpha f(\alpha) \left( \alpha\, K
      - \nabla_i \beta^i \right)
      = \alpha f(\alpha)\,6\,\partial_t\phi\,,
\label{c1pluslog}
\end{equation}
where $f$ is an arbitrary positive function of $\alpha$. The second equality in Eq.~(\ref{c1pluslog})
is due to the evolution equation for $\phi$, namely
\begin{equation}
6\,\partial_t\phi =  -\alpha\, K + \nabla_i \beta^i \,.
\end{equation}
Alcubierre et al.\,\,\cite{Alcubierre2003c} showed how
this family of slicing conditions can be considered a generalization of
a fixed densitized lapse.
For our purposes, it will be sufficient to reformulate Eq.~(\ref{c1pluslog})
in terms of the densitized lapse. This leads to
\begin{equation}
  \partial_t q = q\,(f-n)\, 6\,\partial_t\phi\,. \label{d1pluslog}
\end{equation}
The 1+log slicing is obtained for the special choice of
\begin{equation}
  f=\alpha^{-1} =q^{-1}\,e^{-6n\phi}\,. \label{1+LOG}
\end{equation}

\section{Numerical Results}
\label{RESULTS}
For the 3-dimensional simulations presented here, we evolve the
data according to the BSSN equations
[Eqs.\,(6)-(10) of Paper I]
and actively enforce $\hat{A}^{i}\,_{i}= 0$ after each time
step. We always use a Courant factor of $0.25$.
The other grid parameters as well as gauge and boundary conditions will
be discussed separately for each case studied below.

\subsection{Evolutions of a single static black hole}
\begin{figure}[t]
  \includegraphics[width=200pt, height=250pt, angle=-90]{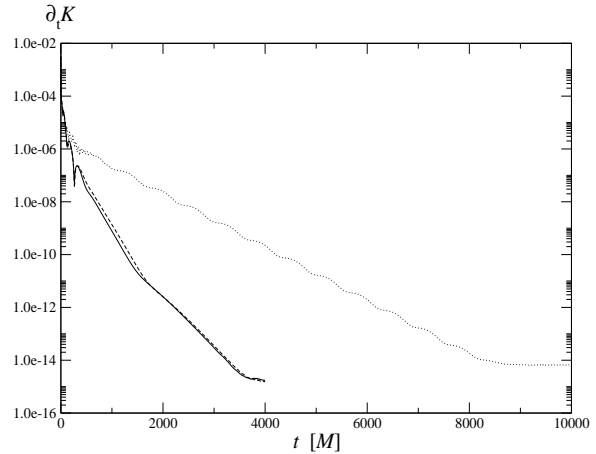}
  \caption{The $\ell_2$-norm of the time derivative of $K$
           for the evolution of a static iEF black hole in
           octant symmetry using 1+log slicing expressed in terms
           of the lapse (solid curve) and densitized lapse (dashed curve).
           The dotted line is from a simulation using an algebraic
           densitized lapse}
  \label{IEFSTABLE}
\end{figure}

%
%
We first consider the numerical evolution of spacetimes containing a single
static black hole. As mentioned above, this problem can be considered
essentially solved in non-linear 3+1 evolutions
\cite{Alcubierre2001b, Yo2002}. It is imperative, therefore, that
any modification of the code preserves the capability of accurately
evolving this scenario with long-term stability. 
For the evolutions presented here, 
the excision region has cubical shape within a radius of $1.5\,M$.
The computational domain 
assumes octant symmetry and has size $12\,M$. 
Fig.\,\ref{IEFSTABLE} shows the 
$\ell_2$-norm of the time derivative
of $K$ for three simulations. The results show that,
as expected for static solutions, $\partial_t K \rightarrow 0$.

The solid and dashed lines are from runs with resolution $2\,M/5$ 
in which simple excision is used \cite{Alcubierre2001b}. That is,
the grid-functions on the excision boundary are updated
by copying the time derivative from the exterior neighbor closest 
to the normal direction. For the dotted line, 
we follow Paper I and use instead a second-order
extrapolation of the updated grid variables on the excision
boundary instead of merely copying the time derivative. The
resolution for this run is $M/5$

Besides the way updating at the excision boundary is handled, 
these three runs differ in the recipe for fixing the slicing. 
For the solid line, we use the classical 1+log condition in terms of 
$\alpha$; that is, the lapse is computed from Eq.~(\ref{c1pluslog}). 
In the dashed line, we apply 
instead the densitized 1+log version given by Eq.~(\ref{d1pluslog}).
It is then clear in Fig.~\ref{IEFSTABLE} that for simple excision
the choice of classical versus densitized
1+log slicing condition does not affect the quality of the simulation. 

The run represented by the dotted line was intended to
address the question of whether such long-term stable evolutions
can also be achieved if we use the purely algebraic
gauge conditions given by Eq.~(\ref{DEFQ}),
with $q$ obtained from the exact iEF solution. The answer is affirmative;
that is, condition (\ref{DEFQ}) in conjunction with second-order
extrapolation on the excision boundary yields
qualitatively similar results to those depicted by the solid and dashed lines.
The motivation for switching for this run from simple excision to 
an excision with extrapolation of the grid-functions
lies in anticipation of 
the wobbling black hole simulations. When the evolved black hole solution
is time-dependent (as is the case of a wobbling black hole), simple excision
is no longer suitable. The copying of time-derivatives at the excision
boundary used in simple excision becomes
effectively a boundary condition on the spatial derivatives of grid-functions.
This boundary condition is 
incompatible with the outflow nature of the excision boundary. 

Comparison of the dotted line with the solid and dashed lines in
Fig.~\ref{IEFSTABLE} shows that the algebraic densitized lapse run
possesses long-term stability. The difference is mostly in the variation
of the time when machine precision is reached. In summary,
using an analytic densitized lapse $q$ or equivalently 
an algebraic lapse $\alpha$,
we are able to evolve a single black hole with stability properties
comparable to those obtained with differential conditions such as
1+log slicing and the $\Gamma$-driver condition for the shift vector.
While it is not clear how helpful such analytic slicing conditions will be for
simulations of a merging binary black hole, the introduction of a
densitized lapse makes them at least available for serious consideration
in long-term stable simulations carried out with the BSSN scheme.

\subsection{Evolutions of single moving black holes}
%
%
We now turn our attention to single black holes moving across the
numerical grid. One way to obtain such a scenario is to evolve a single
boosted black hole. The ensuing motion will, however, 
move the black hole off the computational domain.
With the sizes of the computational domain currently restricted by
available hardware resources, this will happen on time
scales significantly shorter than those considered relevant for
black hole orbits or for testing long-term stability of simulations,
namely simulations lasting more than $1000 M$. 

In Paper I, we introduced therefore an alternative approach which facilitates
the motion of single black holes with 
trajectories confined to the computational domain. We transformed the
iEF black hole solution to coordinates $\bar x^\alpha$ such that
\begin{eqnarray}
  t &=& \bar t, \\[10pt]
  x^{i} &=& \bar x^i + \xi^i(\bar t)\,. \label{XPRIME}
\end{eqnarray}
In terms of the new coordinates,
the line element becomes
\begin{equation}
  ds^2 = -\alpha^2 d\bar t^2 + \gamma_{ij}
          \left( d\bar x^{i} + \bar \beta^i\, d\bar t\right) 
          \left( d\bar x^{j} + \bar \beta^j\, d\bar t\right) \,,
\end{equation}
where
\begin{equation}
  \bar \beta^i = \beta^i + \dot\xi^i. \label{BETAPRIME}
\end{equation}

In Paper I, this method was used to move black holes on circular
and bouncing trajectories. Using either a cubical or spherical excision
region, these runs lasted for about $130\,M$, though the apparent horizon
started to intersect the excision region
at about $90\,M$ and the apparent horizon
finder failed to give reasonable results afterwards. The fixed
gauge conditions used in these evolutions were considered a crucial
limiting factor in those evolutions.
 
\begin{figure}[t]
  \includegraphics[height=250pt,width=200pt]{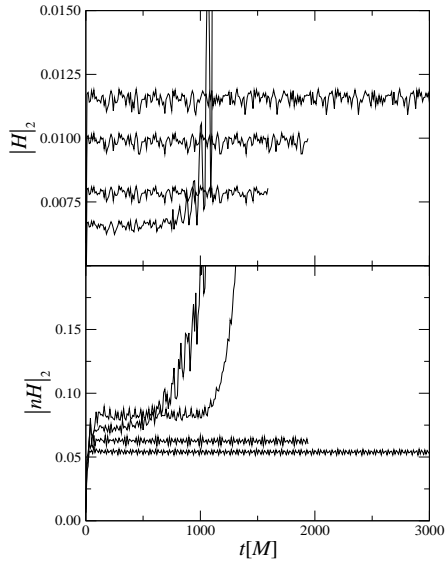}
  \caption{The $\ell_2$-norm of the Hamiltonian constraint (upper panel)
           and normalized Hamiltonian constraint (lower panel)
           are shown as a function of time for the circling black holes for 
           different sizes of the computational domain. See text for details.}
  \label{HAM}
\end{figure}

Using the densitized lapse condition (\ref{DEFQ}) with $n=1$, 
we performed similar simulations to those in Paper I.
We fix the value of $q$ in Eq.~(\ref{DEFQ}) from the exact
single iEF black hole solution transformed according to
Eqs.\,(\ref{XPRIME}), (\ref{BETAPRIME}). 
In Fig.\,\ref{HAM}, we plot the
$\ell_2$-norm of the Hamiltonian constraint (upper panel) and normalized
Hamiltonian constraint (lower panel)
for simulations of a black hole on a circular path with radius $2\,M$ and orbital
frequency $1/(2\pi)$. Equatorial symmetry is assumed for all the
simulations, and the outer boundary conditions consist of setting the
values of the grid-functions to the exact analytic solution.

Fig.~\ref{HAM} depicts results from four simulations. In the upper panel
from top to bottom, each data set corresponds to computational domains
$20\times20\times7\,M^3$,
$20\times20\times10\,M^3$, 
$30\times30\times7\,M^3$ and
$30\times30\times10.5\,M^3$, respectively.
In the lower panel, the correspondence is reversed.
That is, from top to bottom at
early times, each data set is for domains 
$30\times30\times10.5\,M^3$, 
$30\times30\times7\,M^3$, 
$20\times20\times10\,M^3$ and
$20\times20\times7\,M^3$, respectively.
The shortest scale is along the direction perpendicular
to the orbital plane. 

The results in Fig.~\ref{HAM} from the $20\times20\times7\,M^3$ run
should be compared with 
Fig.\,9 of Paper I where simulations lasted $< 130\,M$. 
This comparison clearly demonstrates the tremendous improvement on
the duration of the simulations when a
densitized lapse is used. Some of the evolutions are stable at least for
$6000\,M$, when the runs were terminated due to limitations of computational resources. 
Fig.~\ref{HAM} also demonstrates that, even though the durations of the simulations
have been improved by at least an order of magnitude, the simulations 
continue to be affected by boundary effects. 
This is expected since it is well known that
setting boundary conditions to the exact analytic solution is conducive to 
numerical instabilities. 
\begin{figure}[t]
  \includegraphics[width=250pt, height=250pt]{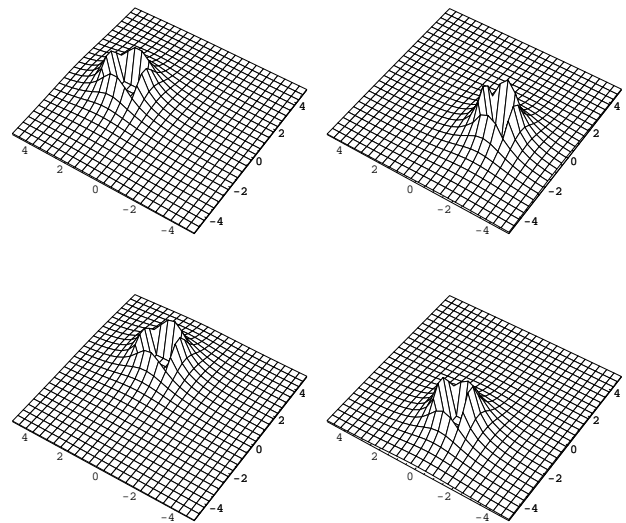}
  \caption{The four panels show snapshots of the evolution of $K$ for a
           circling black hole in the $xy$-plane
           at times $0$, $2000\,M$, $4000\,M$
           and $6000\,M$.  }
  \label{SNAPSHOTS}
\end{figure}
\begin{figure}[t]
  \includegraphics[width=200pt, height=250pt, angle=-90]{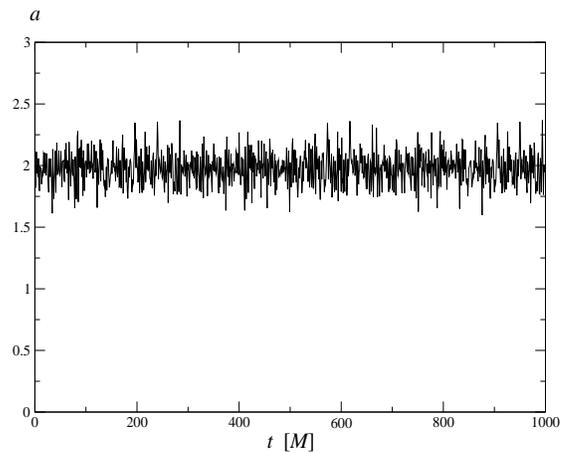}
  \caption{The order of convergence for the circling black hole
           obtained from evolutions using a resolution of $M/4$
           and $M/5$. Second-order convergence is clearly maintained
           for long times.}
  \label{EXPONENT}
\end{figure}

In Fig.\,\ref{SNAPSHOTS} we show snapshots of the evolution obtained
on the domain $20\times 20\times 7\,M^3$ at
$t=0$, $2000\,M$, $4000\,M$, $6000\,M$.
Here we plot the variable $K$ on
the $xy$-plane at $z=0$, namely the orbital plane. 
The excision region is clearly visible
and has been checked lie within the apparent horizon.
We find the apparent horizon area to remain within a few per cent
of its analytic value $16\,\pi\,M^2$ and the deviation from
spherical shape to be of the same order. 
The method we have used to enable the excision region to follow the
motion of the black hole is based on an analysis of the asymmetry and
fall-off behavior of grid variables
near the excision region (see next section for details). This
method is in general not coordinate invariant and its performance needs
to be monitored by verifying that the excision region always remains inside
the apparent horizon. We have verified this for all runs discussed in this
paper \cite{Denzitized_movies}. 

By construction, our numerical code is second-order accurate. 
Fig.\,\ref{EXPONENT} shows the ability of our code to maintain second-order convergence.
Because demonstrating long-term convergence requires considerable
computational resources, we have only 
performed the convergence analysis for times $< 1000\,M$.
The jaggedness in these plots is due to the requirement that the
center of the excision region always be located on a grid point. The excision
mask will therefore move across the computational domain in a discontinuous
fashion.

\begin{figure}[t]
  \includegraphics[width=200pt, height=250pt, angle=-90]{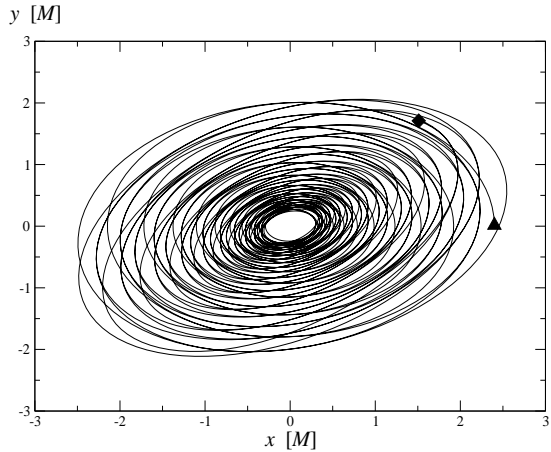}
  \caption{The trajectory of a black hole in-spiraling between radius
           $2.5$ and $0.25$. The initial position is marked by the
           filled triangle and the end position by the diamond.
           The $z$-position will remain zero throughout the
           evolution.}
  \label{TRAJECTORY}
\end{figure}
\begin{figure}[t]
  \includegraphics[width=200pt, height=250pt, angle=-90]{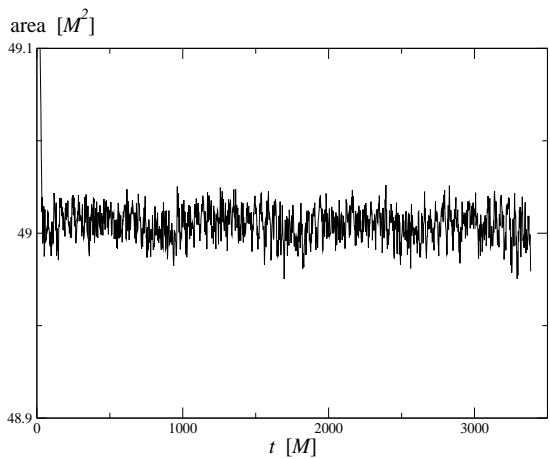}
  \caption{The apparent horizon area for the
           evolution of a spiraling black hole. The value predicted from
           the exact analytic solution is $16\,\pi\,M^2 \simeq 50.27\,M^2$}
  \label{OSCILLATINGAH}
\end{figure}
In order to further test the robustness of our excision infrastructure,
we have performed the same type of evolution using a different
black hole trajectory.
Having in mind the eventual target of simulating in-spiraling binary
black holes, it will be of particular interest to see whether the
code is able to evolve a black hole on an in-spiraling trajectory. We
consider such a scenario by 
fixing the location of the black hole singularity, $\vec{\xi}=(\xi^1,\xi^2)$
[cf.\,Eq.\,(\ref{XPRIME})],
from the solution of the differential
equation
\begin{equation}
  \partial_{tt} \vec \xi + D\,
       \partial_t \vec \xi + \nabla \Phi
       = 0\,,
\end{equation}
with $\xi^3=0$ because the black hole orbit is confined to the plane $z=0$.
The potential is given by $\Phi = C\,[(\xi^1)^2+(\xi^2)^2]$
and $D$ and $C$ are constants. 

We emphasize that none of these quantities has any connection to realistic
astrophysical simulations. They merely serve to
determine the components $\xi^i$ in the coordinate transformation
(\ref{XPRIME}) and, thus, the artificial black hole trajectory. We obtain
an in-spiraling trajectory by choosing $D=C=0.01$.
The black hole is initially placed at a distance of $2.4\,M$ from the
origin at $z=0$ and given an initial purely tangential velocity $v=0.3$. Because
of the damping, the velocity and radius will decrease and the black
hole would eventually approach a steady state at the origin. In order to keep
the evolution dynamic for long times, we switch the sign of the
damping term when the radius shrinks below $0.25\,M$ after which the hole
starts spiraling outward. The damping constant is switched back to a
positive value once a radius of $2.5\,M$ is reached and so on. The
resulting trajectory in the $xy$-plane is displayed
in Fig.\,\ref{TRAJECTORY}. This path corresponds to an evolution lasting
$3500\,M$ when we decided to terminate the evolution.

The apparent horizon finder had no difficulty calculating the
outermost trapped surface every $20\,M$. The resulting horizon area is shown
as a function of time in Fig.~\ref{OSCILLATINGAH}. The area remains constant with
good accuracy and deviates by a few per cent from the analytic value
of $16\pi\,M^2$. Similarly we found the error in the
Hamiltonian constraint normalized with respect to its individual addends
to be constant to within about $6\,\%$.

\section{The excision tracking method}
\label{GAUSSTRACKING}
For the dynamic evolutions of the previous section, we have used a simple
mechanism by which the excision region tracks the movement of the black
hole. We will describe this method in more detail now.
 
In order to make use of the causal disconnection
of the black hole interior it is imperative that the excision region be
located within the event horizon of the black hole. Because the location
of the event horizon requires analysis of the global (in space and time)
data and can thus not be carried out during the evolution itself,
one commonly resorts to searching for apparent horizons. An apparent
horizon (also called a marginally trapped outer surface) is defined
locally and can thus be calculated at each time step (see for
example \cite{Thornburg1996}). The
existence of an apparent horizon guarantees that there is a surrounding
event horizon and thus a causal boundary. 
Ideally one would perform a search for an apparent horizon at each
time step and adjust the location of the excision region correspondingly.
Unfortunately the calculation
of apparent horizons is computationally expensive although recent work
by Thornburg \cite{Thornburg2003} seems to overcome some of these 
difficulties.
It appears, nonetheless, desirable 
to have available a reliable, computationally inexpensive method of
dynamically adjusting the location of the excision regions to
accommodate the movement
of the black holes. One may then calculate the apparent horizon at regular
intervals to monitor the causal consistency of the excision procedure.
\begin{figure}[t]
  \includegraphics[angle=-90]{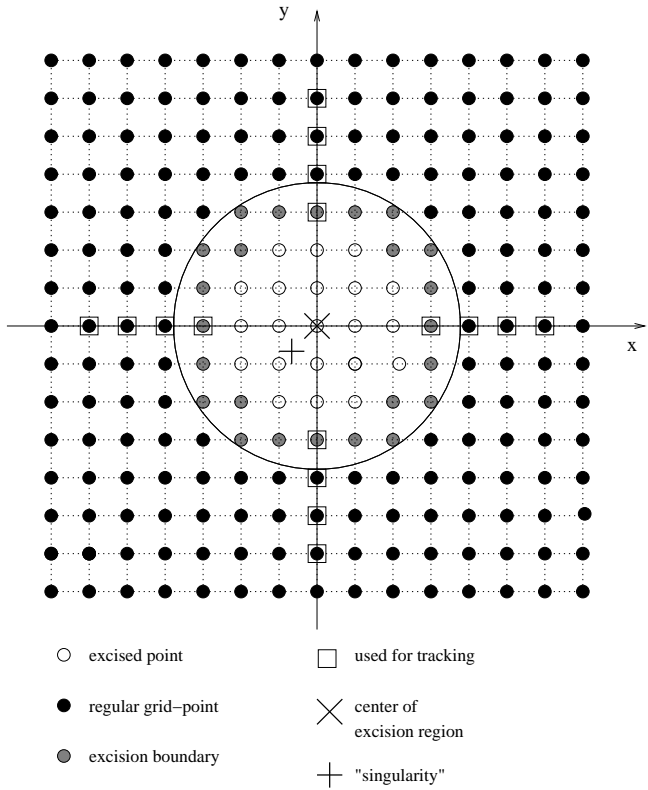}
  \caption{A schematic illustration of the excision region
           (the $z$-direction is suppressed). The $x$ and
           $y$-axes in this diagram are coordinate lines through the current
           excision center ($\times$) whereas the black hole is centered
           at the $+$. Data at at the points marked by a box are used
           to analyze the symmetry properties of a grid variable and
           update the excision center.}
  \label{EXCISION}
\end{figure}
For this purpose we have used a simple scheme which analyzes the
fall-off and symmetry behavior of
evolution variables near the black hole singularity
to predict a reasonable value for the central position of the excision
region. The excision technique used in our Maya code has been described in
detail in Paper I, and we will focus here on those parts
added for the tracking of the excision center.
In Fig.\,\ref{EXCISION} we have schematically plotted a 2-dimensional
slice of a spherical excision region (the precise shape of that region
is irrelevant for our method). The current excision center
is marked by the symbol $\times$. Assuming that the black hole
is slightly offset with respect to this position, we have marked
by a $+$ the desirable location of the excision center. In a loose
sense we might call this the current black hole position. We note,
however, that a
precise definition of a black hole center, as for example in terms of a point-like
singularity, can only be given for a restricted subclass of black holes. 

\begin{figure}[t]
  \includegraphics[width=200pt, height=250pt, angle=-90]{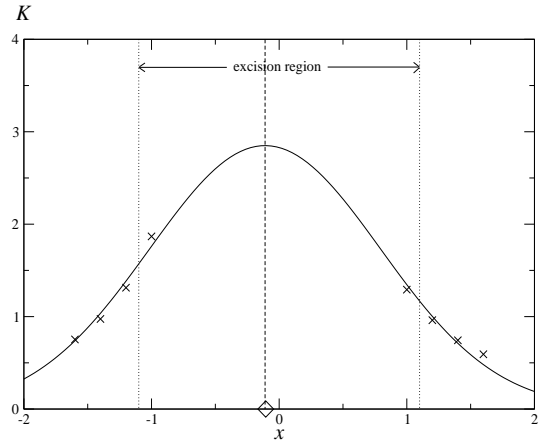}
  \caption{Numerical data ($\times$)
           on the $x$-axis (cf. Fig.\,\ref{EXCISION})
           around the excision region is fitted with a Gaussian curve
           (solid line). The central position of the Gaussian is used
           as the updated $x$ value of the excision center. The
           diamond marks the exact location of the singularity
           used for these data and is well-approximated by the
           fitting algorithm.}
  \label{FIT}
\end{figure}

For concreteness, we now consider analytic data for
a static black hole of mass $M=1$ in iEF coordinates
located at $x=-0.105$, $y=0$, $z=0$.
We describe this data in terms of an excision region with radius $0.7$
centered at $x=0$, $y=0$, $z=0$. Quite obviously this is not the optimal
position of the excision region and the data will show some asymmetry
in their fall-off behavior in the vicinity of the excision boundary.
The idea is then to construct some combination of the field variables which
adequately exhibits this behavior. Such a
combination will in general not be coordinate-invariant and thus
depends on the scenario under investigation.
For all runs presented in this paper, we have found 
the trace of the extrinsic curvature $K$ a perfectly adequate choice.
A 1-dimensional plot in Fig.\,\ref{FIT}
of $K$ along the
$x$-axis through the excision center reveals the asymmetry. The $\times$
symbols in this figure correspond to the values of $K$ at the
8 points marked by boxes on the $x$-axis in Fig.\,\ref{EXCISION}.
At the end of each evolution step in our code we fit a Gaussian
\begin{equation}
   a\,e^{\frac{(x-b)^2}{c^2}}
\end{equation}
through these points. The parameters $a$, $b$ and $c$ are obtained
from $\chi^2$-minimization and the value for $b$ gives us the
$x$-coordinate of the updated excision center. We then proceed
similarly for the $y$ and $z$ direction. If further excision regions are
present in the computational domain they are treated in the same way.
The total number of points used for this method is a free parameter
but we typically find 8 (as in this example) to be sufficient. 

With regard to non-stationary scenarios
it is, of course, possible that a certain asymmetry of the data
might arise from reasons unrelated to the black hole location, as for example
in the case of boosted black holes. We emphasize, however, that the purpose of
this algorithm is not to provide as accurate as possible an estimate
of a black hole center (which in many cases will not be well defined anyway),
but to prescribe a recipe for centering the excision region.
The only
requirement for a healthy evolution is that the excision region be
contained entirely within the apparent horizon.
This may be monitored, for example, by regular calculations of the
apparent horizon. In the numerical evolutions
presented above we have used a few buffer zones (layers of grid points
inside the apparent horizon which are not excised)
and verified that the excision region indeed remains
confined to the interior of the apparent horizon.

\section{Conclusions}
\label{CONCLUSIONS}
We have studied the impact of a densitization of the lapse function
on the stability properties of BSSN evolutions using singularity excision.
We have demonstrated
that the use of a densitized lapse preserves the capability of stably
evolving a single Schwarzschild black hole using the simple excision
technique in combination with a differential slicing condition of the
live 1+log type.
These numerical simulations settle down to a configuration that remains
constant in time up to machine precision. 

We have next considered the evolutions of single static black holes
in the case where the slicing condition 
is a prescribed analytic function of the
spacetime coordinates. In such scenarios numerical codes using the
BSSN scheme have hitherto exhibited significantly poorer stability
properties than those based on hyperbolic formulations. As the use of a
densitized lapse (or a generalized version thereof)
is a necessary ingredient in the majority of hyperbolic
formulations, it has naturally been used in these codes
in place of the lapse function itself. We have bridged the gap
between hyperbolic and BSSN-like formulations,
by evolving a stationary single black hole spacetime using the BSSN
formulation in combination with a densitized version of the lapse function.
We have thus been able to obtain evolutions lasting many thousands
of $M$ in which the time-variations decay to machine precision.
To our knowledge this has so far not even been achieved with
strictly hyperbolic formulations.
Regarding the use of purely analytic gauge conditions for
such scenarios,
our results indicate that the use of a densitized lapse (i.e.\, algebraic
as opposed to fixed slicing)
has been the key advantage of hyperbolic systems over
BSSN-like formulations of Einstein's equations. 
It is not clear to what extent algebraic gauge conditions will
be beneficial in the simulation of in-spiraling binary black holes,
but the results presented here demonstrate how such conditions
can also be used for long-term stable evolutions carried out in the
framework of the BSSN formulation. 

These ideas have also shown advantages in evolving black
holes moving across the computational domain. By virtue of time-dependent
coordinates, we are able to move single iEF black holes through the
computational domain on essentially arbitrary paths. The duration of such
evolutions has previously been limited to about $100\,M$. Because the
motion of the singularity is prescribed in terms of the gauge conditions,
these simulations inevitably used analytic gauge. By reformulating
this in terms of the densitized lapse, we were able to extend the lifetime
of the runs to at least $1000\,M$. This also demonstrates
that the dynamic excision techniques used for our simulations are, in
principle, capable of handling long-term stable evolutions of moving
black holes.
We believe that the ideas presented in this work will be helpful
for the development of long-term stable evolutions of black hole collisions.

\begin{acknowledgments}
Work partially supported by NSF grant PHY-9800973 to Penn State University
and PHY-9800737, PHY-9900672 to Cornell University.
We acknowledge the
support of the Center for Gravitational Wave Physics funded by the
NSF under PHY-0114375.
P.L. and D.S. thank the support while visiting the KITP during which part of 
this work was completed. KITP is supported in part by the NSF under 
PHY-9907949. 
Parallel and I/O infrastructure provided by Cactus. 
\end{acknowledgments}

\end{document}